\documentclass[12pt]{iopart}
\usepackage{graphicx}
\usepackage{bm}

\def\Ang{\AA$^{-1}$}

\def\Xq{$\chi'_{\bf q}$}
\def\Gq{$\Gamma_{\bf q}$}
\def\sqw{$S({\bf q},\omega)$}
\def\Xqw{$\chi''({\bf q},\omega)$}
 
\begin{document}
\jl{03}

\title[Neutron scattering study of YbAgGe]
{Inelastic neutron scattering study of single crystal heavy fermion YbAgGe}
\author{B. F\aa k$^1$,  
D.F. McMorrow$^2$, 
P.G. Niklowitz$^1$, S. Raymond$^1$, E.~Ressouche$^1$, J. Flouquet$^1$,
P.C. Canfield$^3$, S.L. Bud'ko$^3$, Y.~Janssen$^4$, 
and M.J. Gutmann$^5$}

\address{$^1$ Commissariat \`a l'Energie Atomique, 
D\'epartement de Recherche Fondamentale sur la Mati\`ere Condens\'ee, 
SPSMS, 38054 Grenoble, France}
\address{$^2$ Department of Physics and Astronomy,
University College London, Gower Street, London, WC1E 6BT, UK}
\address{$^3$ Ames Laboratory and Department of Physics and Astronomy, Iowa State University, Ames, Iowa 50010, USA}
\address{$^4$ Ames Laboratory, Iowa State University, Ames, Iowa 50010, USA}
\address{$^5$ ISIS Facility, Rutherford Appleton Laboratory, Chilton,
Didcot, Oxon OX11 0QX, Great Britain}

\begin{abstract}
Single crystals of the heavy-fermion compound YbAgGe have been studied by neutron scattering.  The magnetic ordering occurring below $T_1\approx 0.5$ K is characterized by a commensurate propagation vector {\bf k}=(1/3,0,1/3) and the moments in the basal plane of the hexagonal structure. 
The dynamic magnetic susceptibility is dominated by quasielastic spin fluctuations with a characteristic energy $\Gamma_{\bf q}(T)$ of the order of 1 meV. The spins fluctuate predominantly in the basal plane. No spin-wave excitations are observed in the magnetically ordered phase. Below the Kondo temperature, $T_K\sim 20$ K, $\Gamma_{\bf q}(T)$ shows a strong $q$ dependence for wave vectors along the $c^\star$ direction, but is $q$-independent in the basal plane. $\Gamma_{\bf q}(T)$ shows initially a rapid increase with temperature $T$ at the antiferromagnetic zone center, but follows a standard $\sqrt T$ law for other {\bf q} values and for $T>T_K$ in general. These observations classify YbAgGe as a well-behaved heavy-fermion compound with a particular {\bf q}-dependence of the antiferromagnetic spin fluctuations, possibly related to the geometrical frustration of the Yb$^{3+}$ ions. 
\end{abstract}

\vspace{5mm} 20 October 2004
\maketitle

\section{Introduction}
\label{SecIntro}
YbAgGe is a recently discovered Yb-based moderate heavy-fermion (HF) compound 
with a Kondo temperature of $T_K\approx 20$ K 
and a linear coefficient of the specific heat, $\gamma=C_p/T$,
of 570 mJK$^{-2}$mole$^{-1}$ at $T=1.6$ K 
and 150 mJK$^{-2}$mole$^{-1}$ in the zero-temperature limit \cite{Morosan04,Budko04,Katoh04}. 
At low temperatures, YbAgGe shows two magnetic phase transitions,
a first-order transition at $T_1\approx 0.65$ K
and a second-order transition at $T_2=$ 0.8--1.0 K \cite{Morosan04,Umeo04}. 
The small entropy at low temperatures, only 5\% of $R\ln 2$ at $T=1$ 
K, suggests a small-moment ordering \cite{Morosan04}. 
Both magnetic phases disappear with the application of a magnetic 
field of the order of 1--3 and 4--8 T, depending on the field 
orientation \cite{Budko04,Umeo04}. At the same time, non-Fermi liquid behavior is 
observed, with a logarithmic term in the specific heat, $C_p/T\propto 
-\ln T$, and a linear resistivity, $\rho-\rho_0\propto T$, for fields
of the order of 8--12 T \cite{Budko04,Budko05}. 
YbAgGe provides thus an excellent opportunity to study a field-induced 
non-Fermi liquid of a stoichiometric compound in the vicinity of a 
quantum phase transition (QPT). 
We have performed neutron scattering studies of single-crystalline 
samples in the zero-field heavy-fermion regime, which is 
characterized by both the Kondo effect and geometrical frustration. 
The measured dynamic magnetic susceptibility has an energy and 
temperature dependence that is typical for a heavy-fermion compound, 
but the wave vector dependence is unusual. 

\begin{figure} 
\begin{center}
\includegraphics[width=8cm]{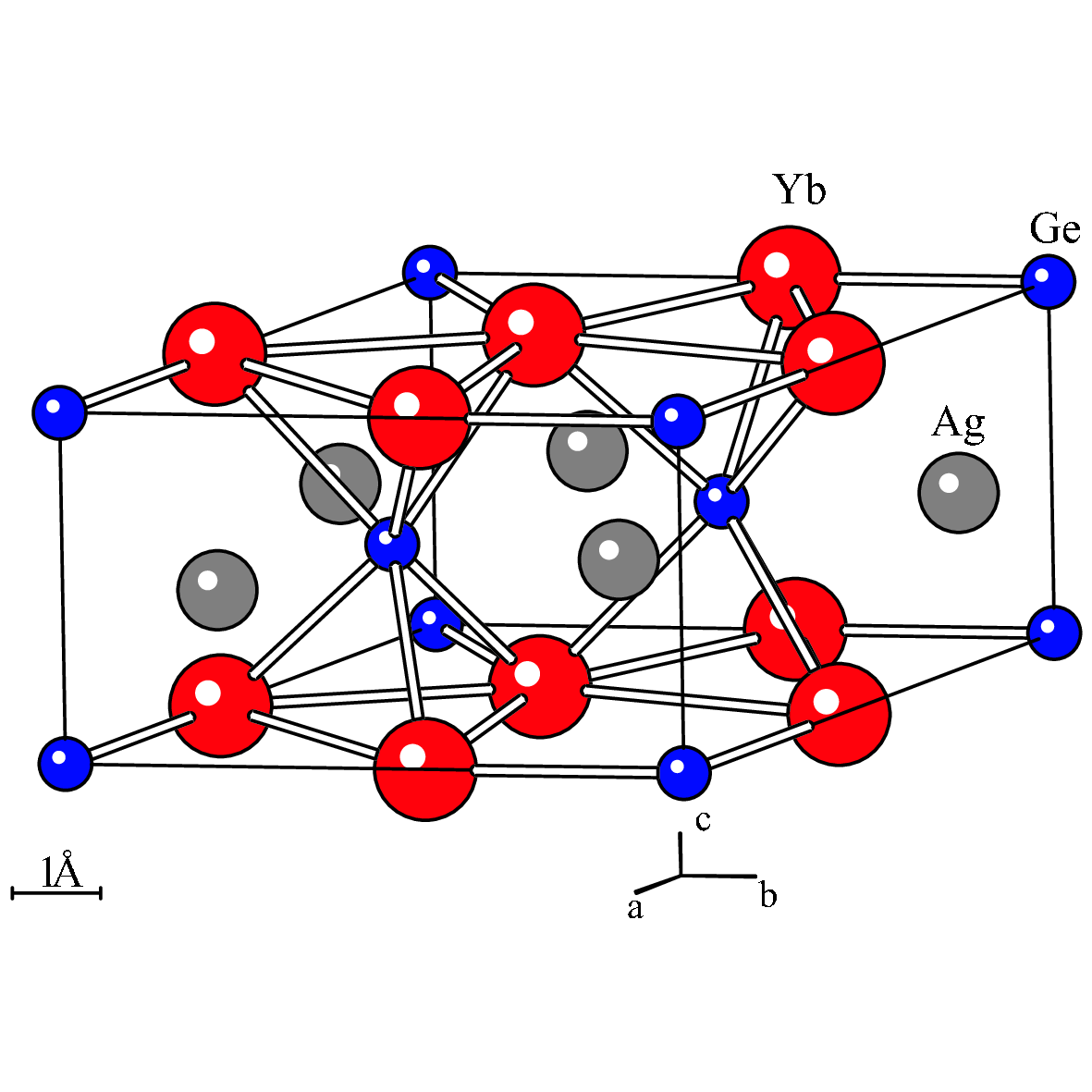}
\end{center}
\caption{Crystallographic structure of YbAgGe.}
\label{FigStructure}
\end{figure}

YbAgGe crystallizes in the hexagonal  ZrNiAl-type crystal structure with the non-centrosymmetric space group $P\bar{6}2m$  and room-temperature lattice parameters of $a=7.05$ and $c=4.14$ \AA\ (see  figure \ref{FigStructure}) \cite{Gibson96}. The Yb$^{3+}$ ions have $4f^{13}$ electronic configuration, which is the hole counterpart to trivalent Ce ($4f^1$) ions. The three Yb ions  per unit cell occupy the same equivalent site in the lattice and lie on a Kagom\'e-like triangular lattice in the basal plane, which for antiferromagnetic coupling leads to geometrical frustration. 
The eight-fold $J=7/2$ multiplet of the Yb$^{3+}$ ion is split into 4 doublets by the crystal electric field. Specific-heat data show a peak at $T=60$ K, which is attributed to the Schottky contribution from the first excited doublet at an energy of $E=9.4$ meV  \cite{Katoh04}. However, the peak height is two times higher than expected for a doublet-doublet excitation, implying that the third and fourth doublets also need to be taken into account. 
The magnetic susceptibility, $\chi$, follows a Curie-Weiss law at high temperatures 
with an effective moment of $\mu_{\rm eff}=4.4\ \mu_B$  \cite{Morosan04,Katoh04},
 i.e. close to the value of the free Yb$^{3+}$ ion, $\mu_{\rm eff}=4.5\ \mu_B$, 
 and a low-temperature easy-plane 
anisotropy of $\chi_{ab}/\chi_c\approx 3$ \cite{Morosan04,Budko04,Katoh04}. 
A small maximum at $T=4$ K is seen in $\chi_a$  \cite{Morosan04,Katoh04}, 
reflecting the onset of antiferromagnetic correlations. 
The corresponding crystal-field scheme, which has the $a$-axis as 
principal axis, has been confirmed by inelastic neutron scattering 
measurements on a polycrystalline sample, which shows a clear peak at 
$E=12$ meV \cite{Matsumura04}. These measurements also revealed Kondo-type spin 
fluctuations with a characteristic energy of $\Gamma=0.9$ meV at low 
$T$ and a $\sqrt T$ behavior at higher $T$. 

We have performed neutron scattering measurements on large single 
crystals of YbAgGe. The sample growth and experimental conditions are 
discussed in section \ref{SecExp}. Section  \ref{SecOrder} presents neutron scattering 
measurements of the magnetic structure of the low-$T$ phase 
below $T_1$. Inelastic neutron scattering measurements of the dynamic 
magnetic susceptibility are presented in section \ref{SecInelastic}  and the results are 
discussed and compared with other $4f$ heavy-fermion materials in section  \ref{SecDisc}.

\section{Experimental}
\label{SecExp}
Two single crystals, each of approximate weight of 2 g, were grown using a Ag-Ge rich self-flux method \cite{Morosan04}. Smaller crystals using the same growth technique have been fully characterized by macroscopic measurements \cite{Morosan04,Budko04}. Neutron scattering measurements were performed on the single-crystal white-beam time-of-flight diffractometer SXD at the ISIS spallation source and on the cold triple-axis spectrometer IN14 of the ILL high-flux reactor.  In this paper, we label the coordinates in reciprocal space using hexagonal reduced lattice units (r.l.u.) with indices ($hkl$), where a point in reciprocal space is given by {\bf Q}$=h{\bf a}^\star+k{\bf b}^\star+l{\bf c}^\star$ with ${\bf a}^\star=4\pi/(\sqrt 3 a)\, \hat{x}$ and ${\bf c}^\star=(2\pi/c)\, \hat{z}$. 

On SXD, one single crystal was aligned with the $c$ axis vertical and glued by black Stycast to a Cu support. The assembly was mounted in a $^3$He refrigerator which was inserted in a standard helium cryostat. Both the cryostat and the $^3$He insert were equipped with Al tails. Measurements were performed for three different orientations of the crystal (rotated around the $c$ axis) at two temperatures: $T=0.32$  K and $T=2$ K. All integer reflections could be indexed in the $P\bar{6}2m$ space group. The magnetic scattering was obtained by subtracting the $T=2$ K data from the low-$T$ data, which removed nuclear Bragg reflections from the sample and powder lines from the cryostat.

On IN14, two single crystals of YbAgGe were aligned together and mounted on a Cu support using Cu straps and Cu mesh for thermal contact. The assembly was mounted in a dilution fridge with ($h0l$) as horizontal scattering plane. IN14 was used in long-chair configuration with a vertically focusing PG(002) monochromator, a horizontally focusing PG(002) analyzer, and horizontal collimations of guide-60'-open-open. Most measurements were performed at fixed final wave vectors of $k_f=1.97$ and $k_f=1.30$ \Ang, with energy resolutions for elastic incoherent scattering of 0.442~(2) and 0.117~(1) meV, respectively. Higher-order harmonics were removed from the scattered beam by pyrolytic graphite or liquid-nitrogen cooled Be filters. 
The intensity of the magnetic scattering so obtained  is proportional to the dynamic structure factor \sqw,
\begin{equation}
S({\bf q},\omega) = {1 \over 1-\exp(-\hbar\omega/k_BT)} \ \chi''({\bf q},\omega),
\label{EqSqw}
\end{equation}
where the dynamic magnetic susceptibility, \Xqw, is obtained from \sqw\ after subtracting a constant background term, determined at negative energy transfers at low temperatures or at positive energy transfers with the analyzer turned off from the Bragg condition by 10 degrees. 
Measurement were performed at temperatures (of the mixing chamber) between $T=0.06$ and 50 K in zero magnetic field. At the lowest temperatures, the sample could have been slightly warmer ($\sim 0.1$ K) than the mixing chamber due to beam heating.

\begin{figure} 
\begin{center}
\includegraphics[width=8cm]{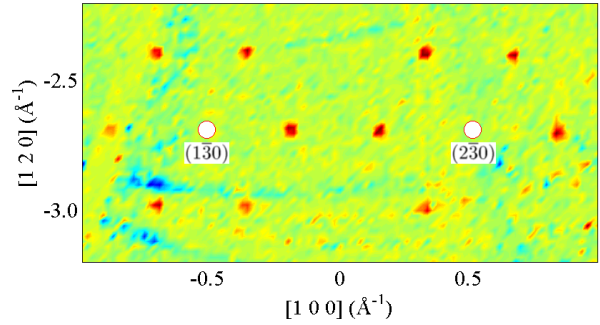}
\end{center}
\caption{Color map of the low-$T$ magnetic scattering in YbAgGe, taken as the difference between data at $T=0.32$ and 2.0 K, in the $(hk)$ scattering plane for $l=1/3$. The six-fold symmetry of the magnetic satellites characterized by a propagation vector of {\bf k}=(1/3,0,1/3) is easily seen.}
\label{FigSXD}
\end{figure}

\section{Magnetic order}
\label{SecOrder}
The magnetic scattering in the low-temperature phase at $T=0.32$ K (after subtraction of the $T=2$ K data) measured on SXD is shown in  figure \ref{FigSXD}. Non-integer superstructure peaks due to magnetic order are clearly observed. They can be indexed with a propagation vector of {\bf k}=(1/3,0,1/3). The intensity of the superstructure peaks decreases with increasing $Q$, as expected from the magnetic form factor. The intensity is weak for small $l$ indices, which suggests that the moments are predominantly in the basal plane of the hexagonal structure, in agreement with the easy-plane anisotropy observed in the magnetic susceptibility measurements.

\begin{figure} 
\begin{center}
\includegraphics[height=5cm]{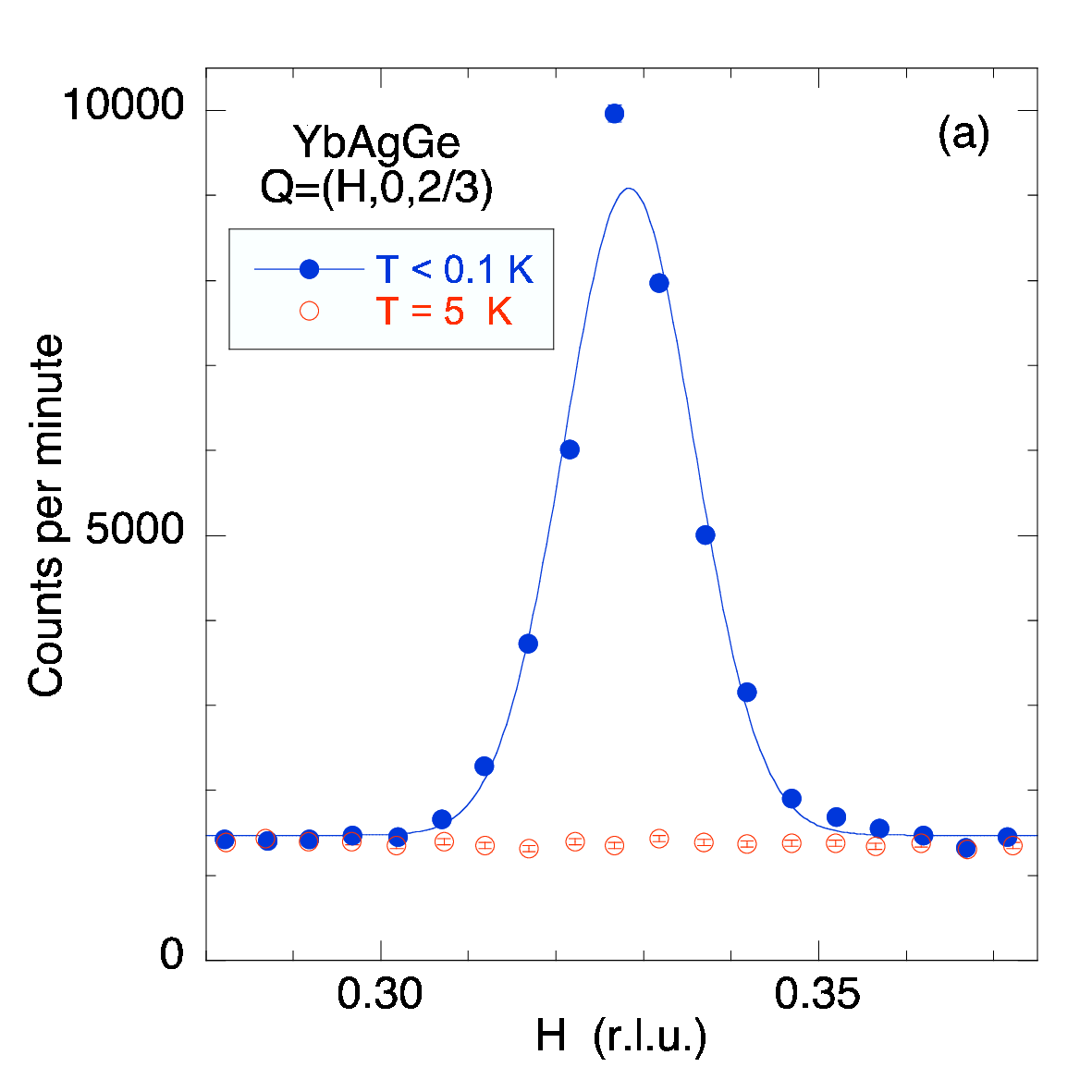}
\includegraphics[height=5cm]{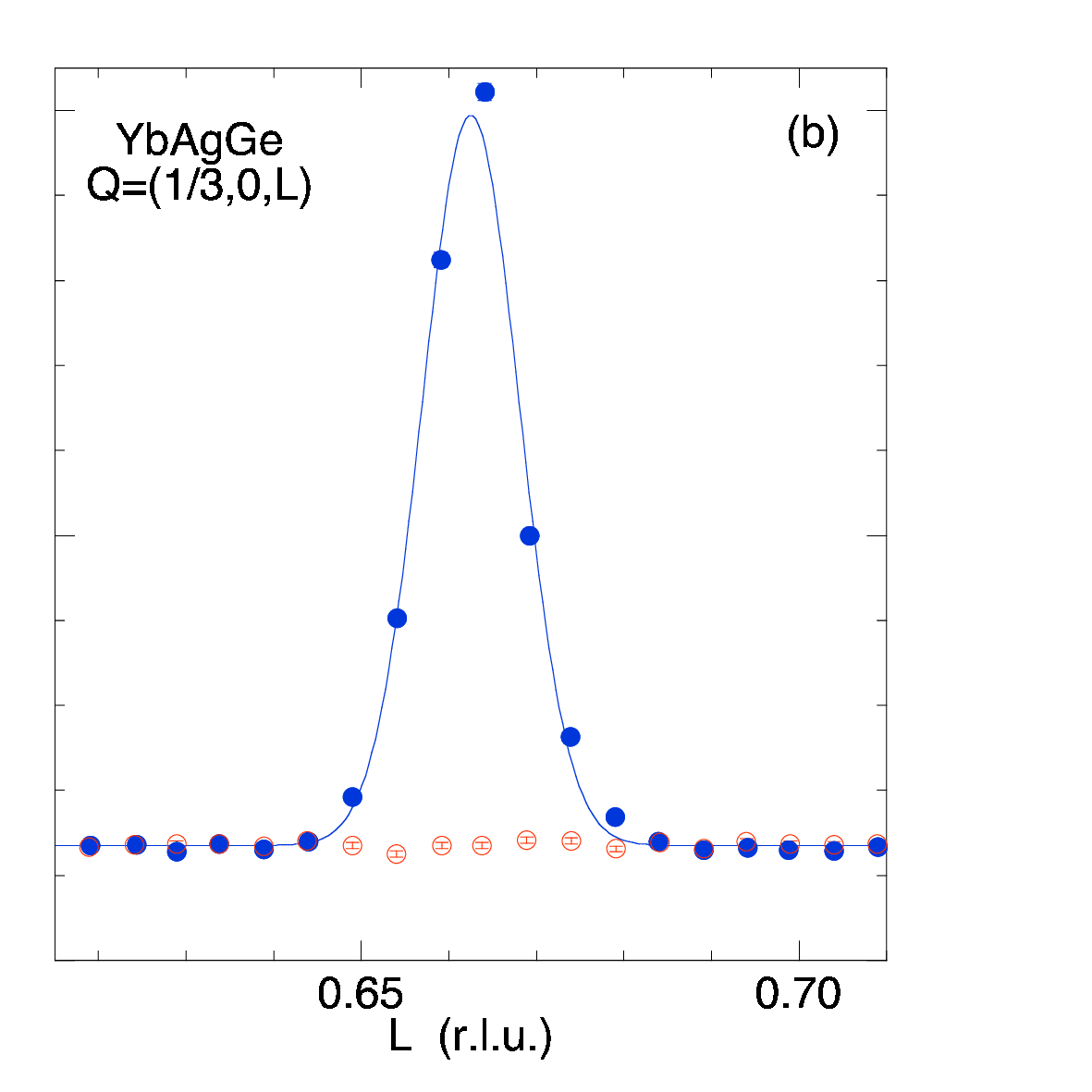}
\end{center}
\caption{Scans at zero energy transfer along the (a) $h$ and (b) $l$ direction of the {\bf Q}=(1/3,0,2/3) magnetic Bragg peak of YbAgGe measured on IN14 using $k_f=1.3$ \Ang. Closed and open circles correspond  to temperatures of 0.1 and 5.0 K, respectively. The lines are Gaussian fits.}
\label{FigIN14Bragg}
\end{figure}

\begin{figure} 
\begin{center}
\includegraphics[width=7cm]{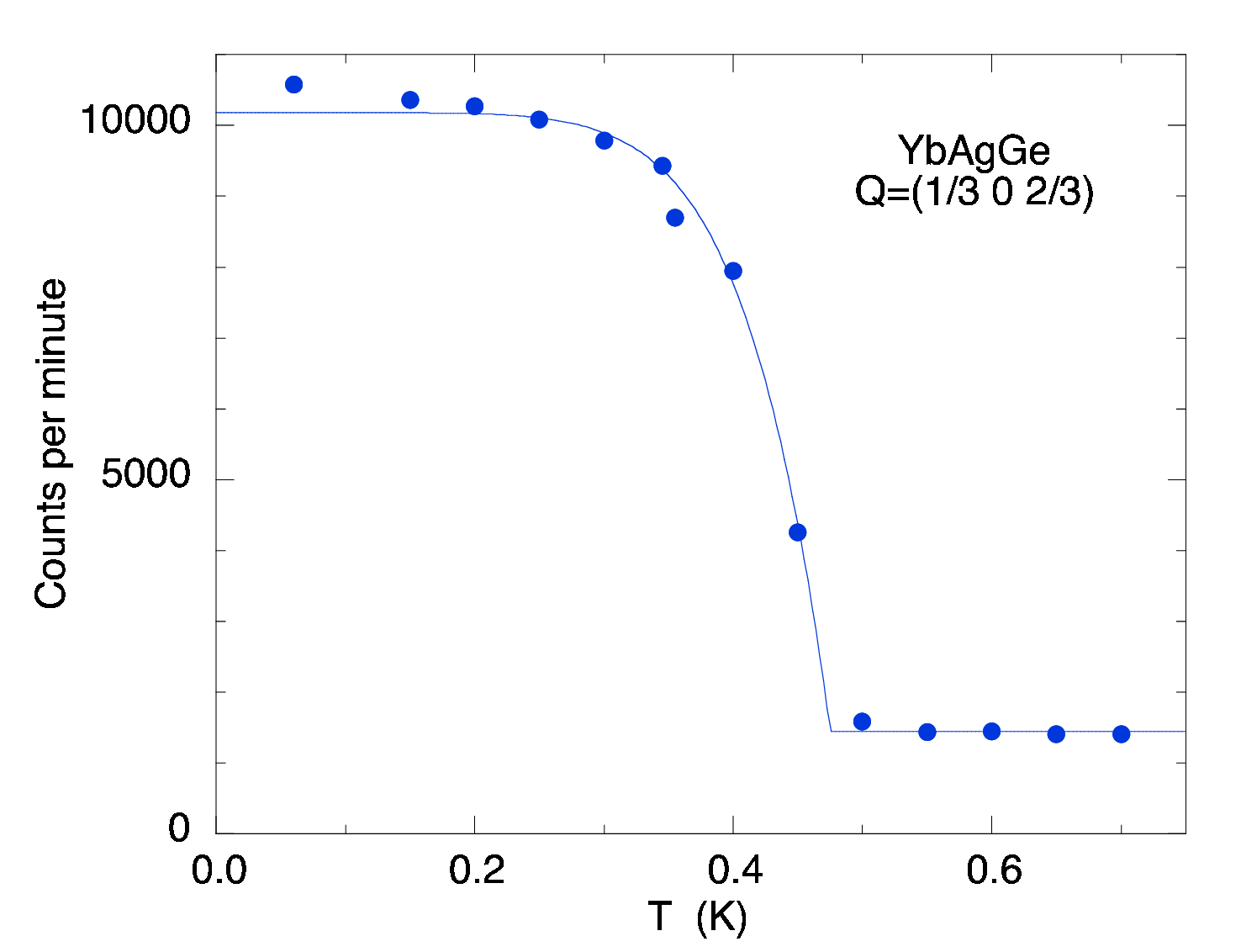}
\end{center}
\caption{Temperature dependence of the magnetic Bragg peak intensity at {\bf Q}$=(1/3,0,2/3)$ of YbAgGe measured on IN14 using $k_f=1.3$ \Ang. The line is a guide to the eye. }
\label{FigM2ofT}
\end{figure}

Scans of the magnetic Bragg peaks along the $h$ and $l$ directions were performed below and above the magnetic ordering temperature $T_1$ on IN14 at $k_f=1.3$ \Ang\ (see  figure \ref{FigIN14Bragg}). These scans confirm the propagation vector of {\bf k}=(1/3,0,1/3) observed on SXD. The magnetic order is long-range, as seen from the resolution-limited width of the  peaks. The temperature dependence of the magnetic Bragg peak intensity is shown in  figure \ref{FigM2ofT}. The transition temperature determined from these measurements is $T_1=0.48$~(2) K, which is similar to 
that from bulk measurements. 
The rapid drop of the magnetic intensity near $T_1$ is reminiscent of a first-order phase transition.
Above $T_1$, no magnetic super-lattice peaks were observed. 
Further measurements are planned to investigate if the phase between $T_1\sim 0.5$ K and $T_2\sim 0.9$ K corresponds to a magnetically ordered phase. 
A full determination of the magnetic structures below $T_1$ is also in progress.

\section{Dynamic magnetic susceptibility}
\label{SecInelastic}
\begin{figure} 
\begin{center}
\includegraphics[width=7cm]{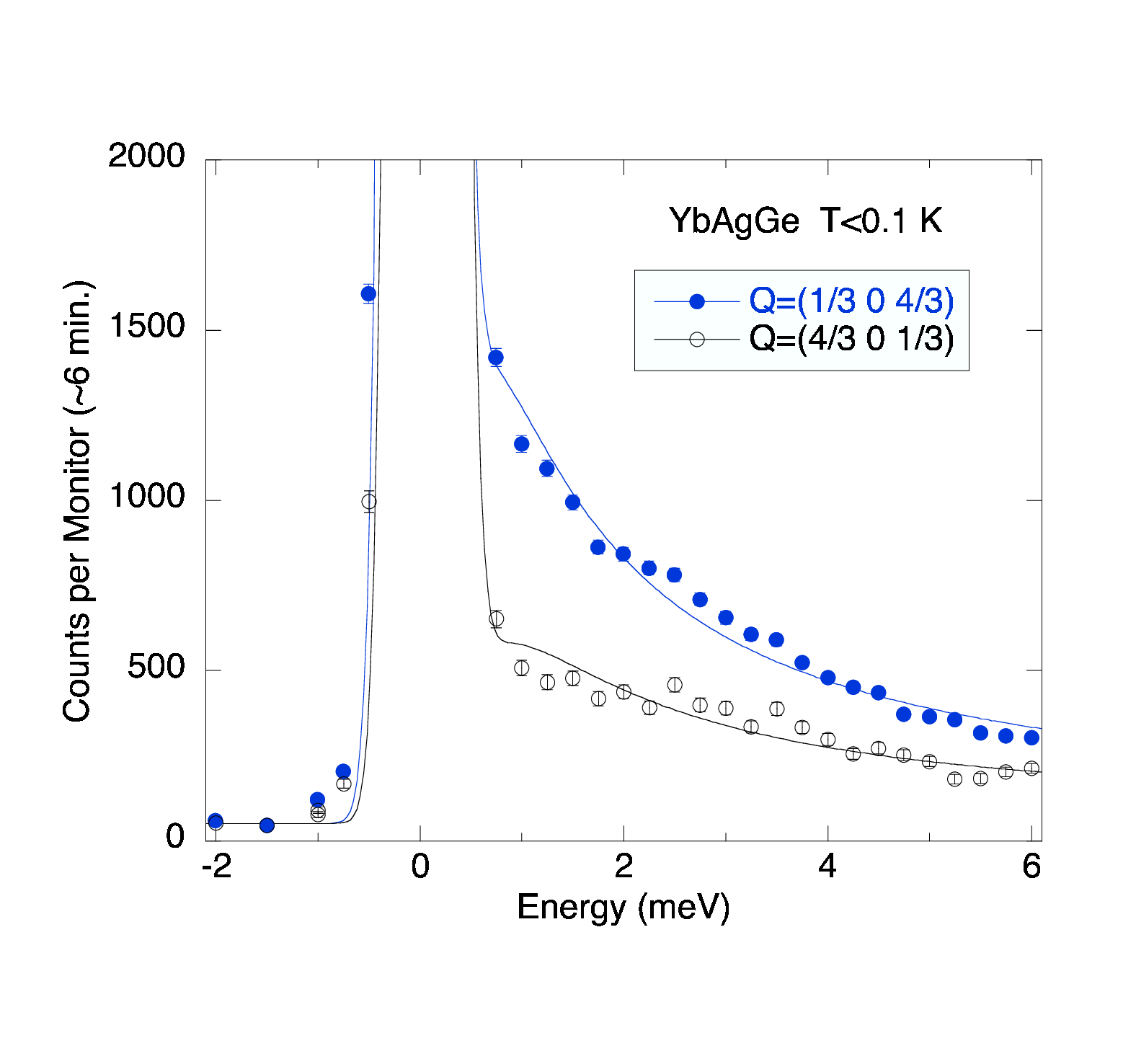}
\end{center}
\caption{Energy scans of the quasielastic magnetic scattering at two equivalent antiferromagnetic positions,  measured on IN14 at $T<0.1$ K using $k_f=1.97$ \Ang. The lines are fits to a quasielastic Lorentzian, Eqs.\ (\ref{EqSqw}-\ref{EqLor}), plus a Gaussian describing the elastic scattering.}
\label{Fig197}
\end{figure}

Our inelastic neutron scattering (INS) measurements on single crystalline YbAgGe show the existence of strong magnetic quasielastic scattering arising  from spin fluctuations. The dynamic magnetic susceptibility \Xqw\ 
is quite well described by a quasielastic Lorentzian,
\begin{equation}
\chi''({\bf q},\omega) = \frac{ \omega\chi'_{\bf q}\Gamma_{\bf q} }{ \omega^2+\Gamma_{\bf q}^2},
\label{EqLor}
\end{equation}
with a characteristic energy of $\Gamma_{\bf q}\sim 1$ meV.  This is in agreement with neutron scattering measurements on a polycrystalline sample by Matsumura \etal \cite{Matsumura04}. 
Matsumura \etal also observed a crystal-field excitation at approximately 12 meV, which is outside the energy window of the present measurements. 
The intensity of the spin-fluctuation scattering at the antiferromagnetic position {\bf Q}=(1/3,0,4/3) is nearly twice as big as at the equivalent position {\bf Q}=(4/3,0,1/3) (Figure \ref{Fig197}). Since INS probes magnetic fluctuations perpendicular to the total wave vector transfer {\bf Q}, this suggests that the spins fluctuate predominantly in the basal plane of the hexagonal structure. This is in agreement with the easy-plane anisotropy in bulk magnetic susceptibility measurements. 

As we will show  in detail below (section \ref{SecTdep}), the quasielastic magnetic scattering is independent of temperature between $T=0.1$ and 0.5 K, i.e. the quasielastic scattering exists in the low-temperature antiferromagnetically ordered phase. This may be a feature common to $4f$ heavy-fermion materials close to a quantum phase transition, as e.g.\ quasielastic magnetic scattering has also been observed in CePd$_2$Si$_2$ \cite{CePd2Si2} and CeIn$_3$ \cite{CeIn3}  in the ordered state. In these compounds, the spin fluctuations coexist with spin waves. In YbAgGe, we have not observed any spin waves. This may be because of the weak magnitude of the ordered moment.

\begin{figure} 
\begin{center}
\includegraphics[width=11cm]{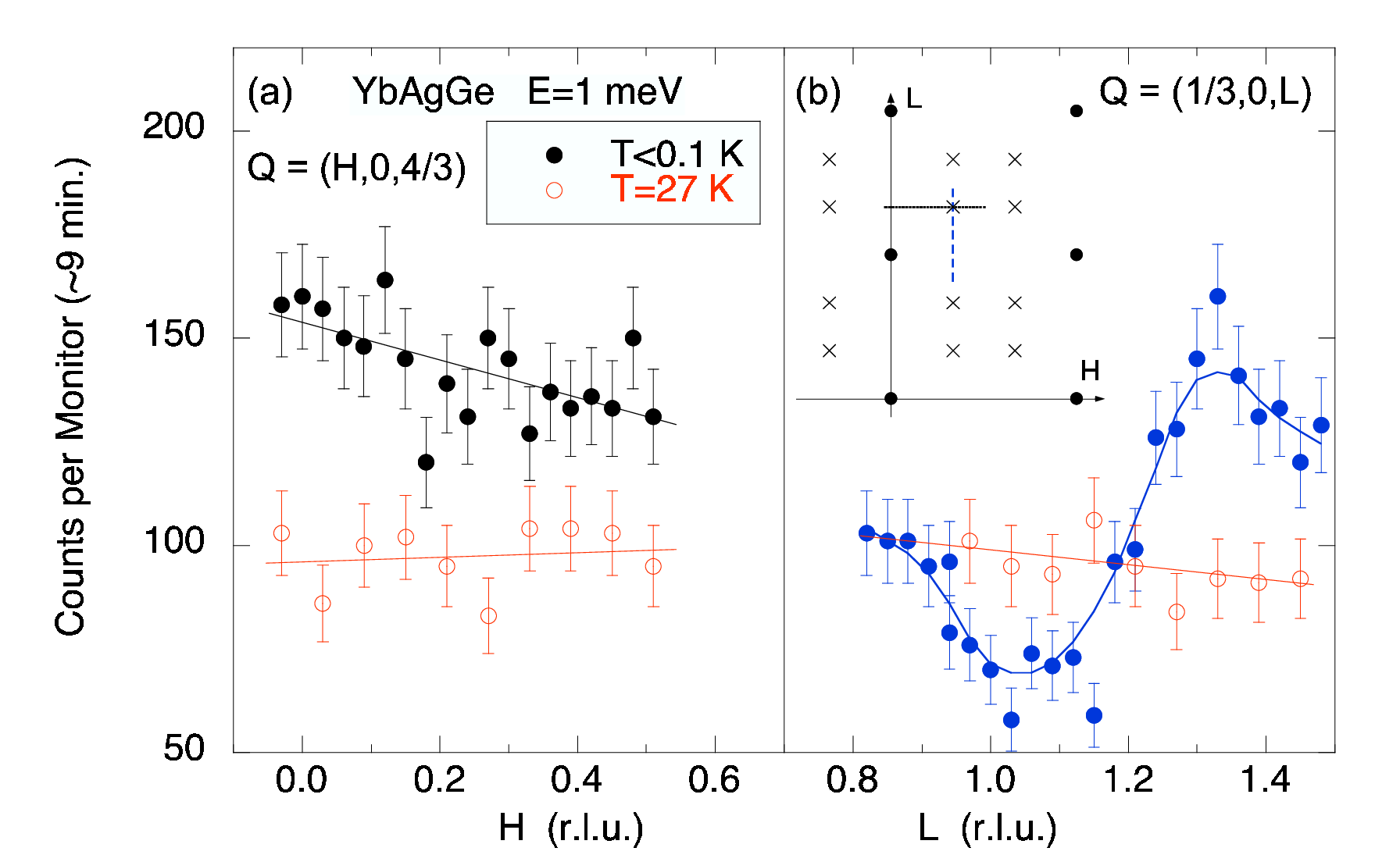}
\end{center}
\caption{Scans in wave vector {\bf Q} along the (a) $h$ and (b) $l$ direction of the spin-fluctuation scattering at an energy transfer of $E=1$ meV, measured on IN14 using $k_f=1.30$ \Ang. Closed and open circles correspond to $T<0.1$ and $T=27$ K, respectively. The lines are guides to the eye. The inset shows the scan directions in the $(h0l)$ plane. }
\label{FigQscan}
\end{figure}

\subsection{{\bf Q} dependence}
\label{SecQdep}

\begin{figure} 
\begin{center}
\includegraphics[width=7cm]{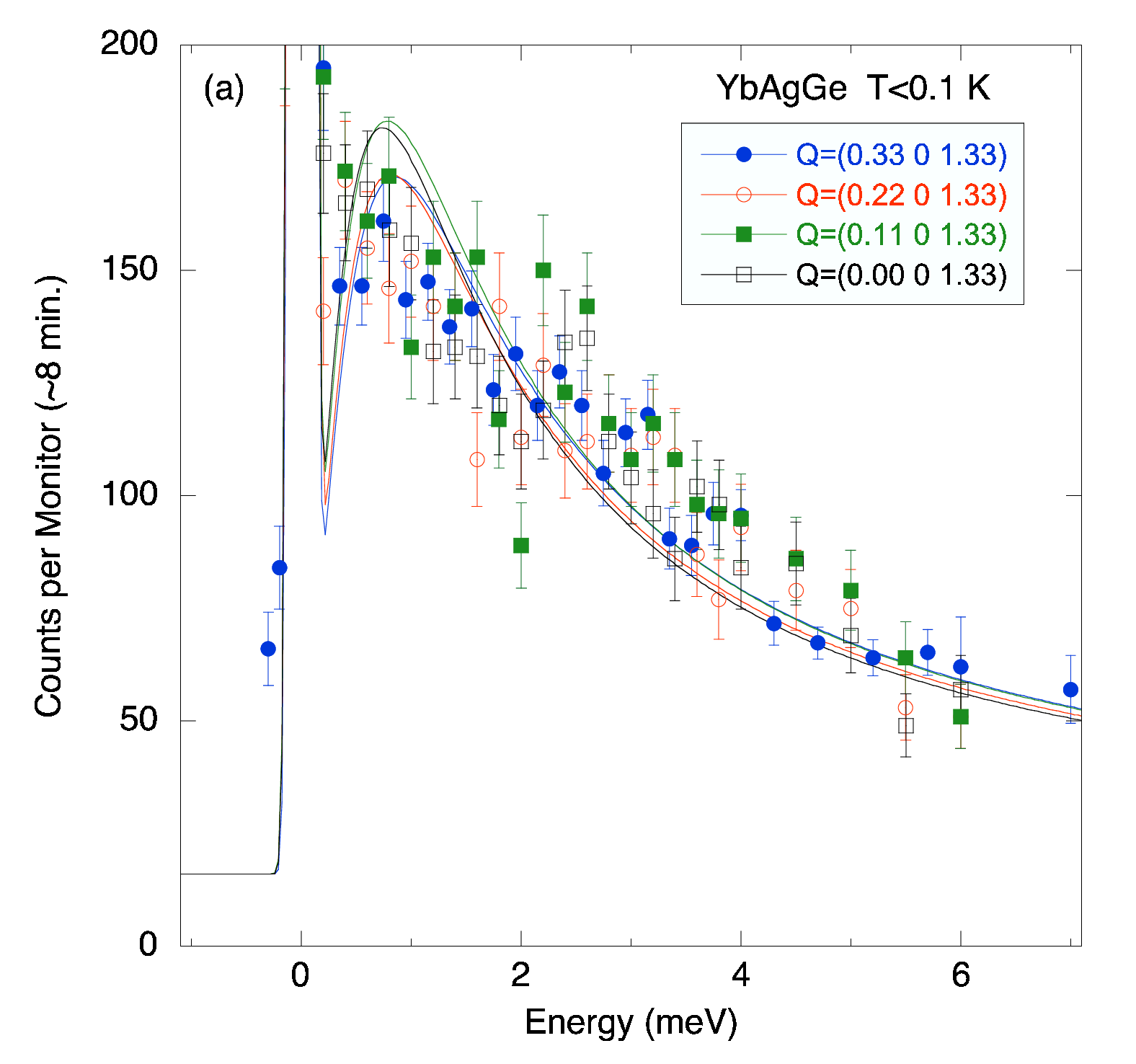}
\includegraphics[width=7cm]{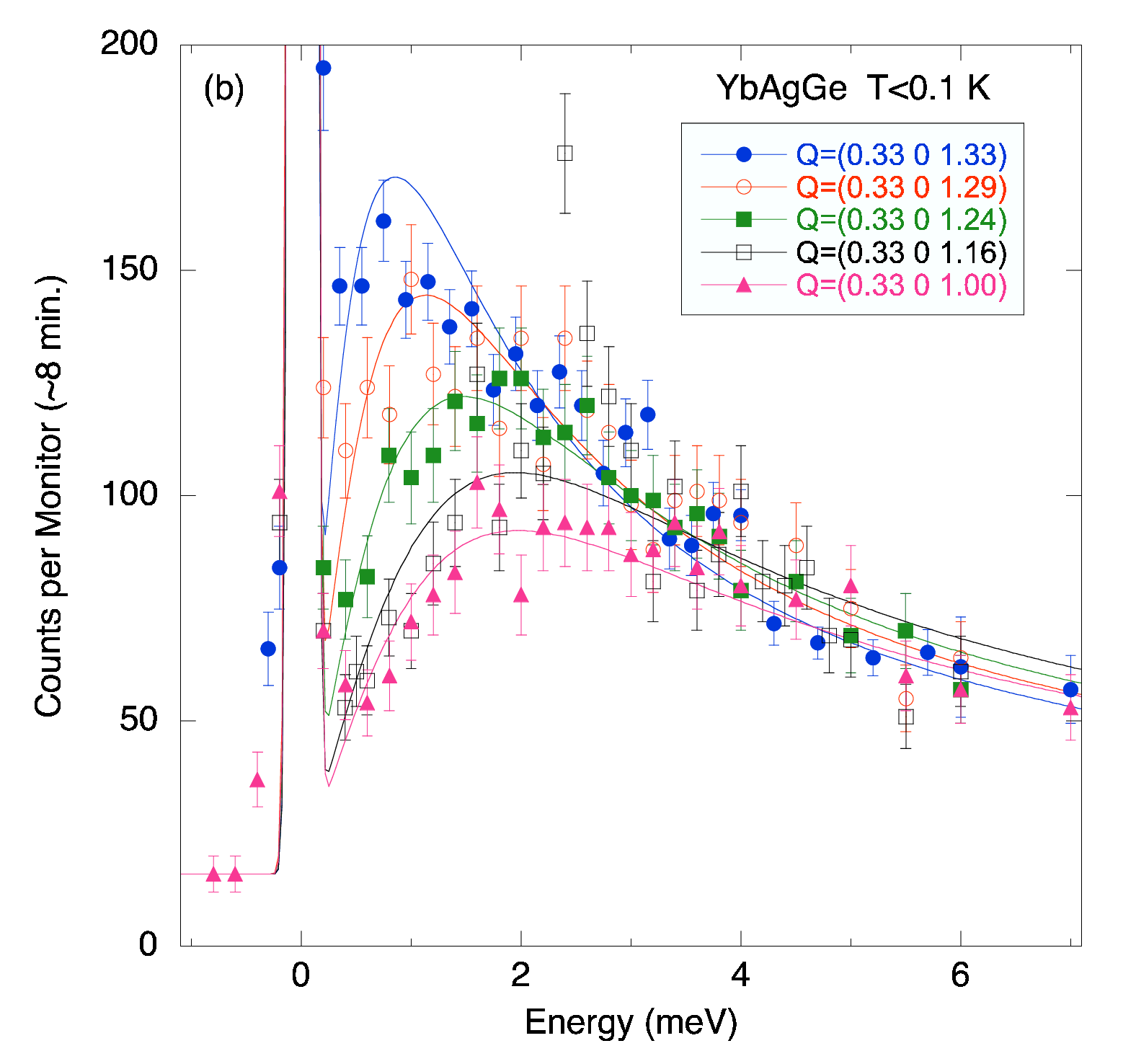}
\end{center}
\caption{Energy scans of the quasielastic magnetic scattering at low temperatures,  $T<0.1$ K, for different {\bf q} values along the  (a) $h$ and (b) $l$ direction, measured on IN14 using $k_f=1.30$ \Ang. The lines are fits to a quasielastic Lorentzian, Eqs.\ (\ref{EqSqw}-\ref{EqLor}), plus a Gaussian describing the elastic scattering.}
\label{FigEscan}
\end{figure}

\begin{figure} 
\begin{center}
\includegraphics[width=9cm]{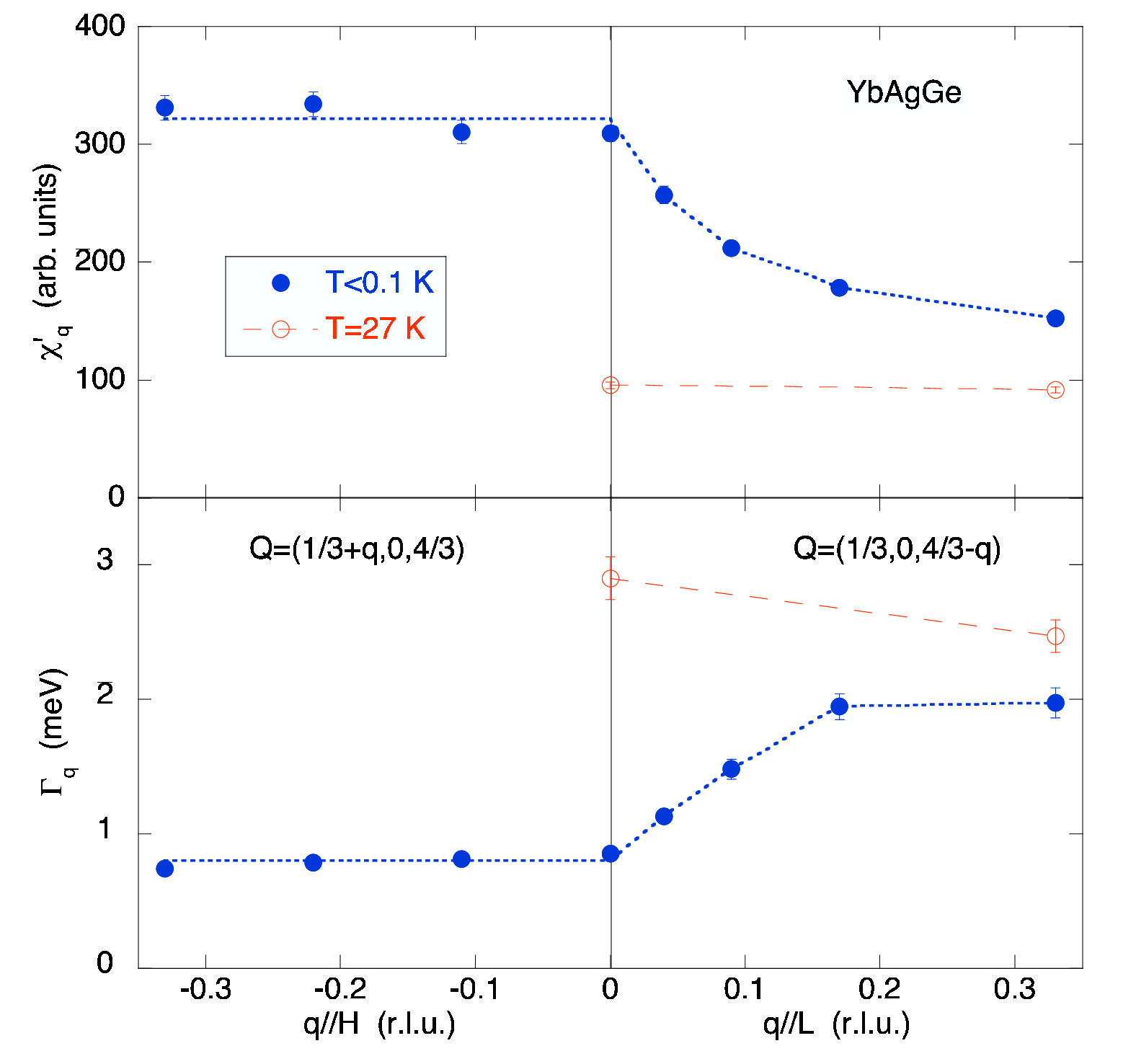}
\end{center}
\caption{Wave vector dependence of the static susceptibility $\chi'_{\bf q}$ and the characteristic energy $\Gamma_{\bf q}$ of the quasielastic magnetic scattering along the $h$ (left part) and $l$ (right part) direction at $T<0.1$ K (closed circles) and $T=27$ K (open circles). The lines are guides to the eye. }
\label{FigQdisp}
\end{figure}

Figure \ref{FigQscan} shows {\bf Q} scans in two different directions taken at an energy of  $E=1$ meV, where the intensity of the spin-fluctuation scattering is maximal. The intensity at low temperatures is clearly modulated in the $l$ direction. This shows that the spin fluctuations are correlated in {\bf Q} and hence does not correspond to localized (or single-site) fluctuations. 
Such correlated spin fluctuations have been observed in other heavy fermion systems. 
The modulation of the intensity is not simply a modulation in $\chi'_{\bf q}$, 
but corresponds also to a {\bf q} dependence of the characteristic energy $\Gamma_{\bf q}$. This is clearly seen in energy scans at different {\bf Q} values along $l$ taken at low temperatures (figure \ref{FigEscan}b). 
On the other hand, the {\bf Q} scan taken in the $h$ direction (cf.\ figure \ref{FigQscan}) is nearly flat, and energy scans along the $h$ direction (cf.\ figure \ref{FigEscan}a) are nearly the same. The corresponding ``dispersion'' of the characteristic energy $\Gamma_{\bf q}$ as well as the {\bf q}-dependent susceptibility $\chi'_{\bf q}$, obtained from fits of Eqs.\ (\ref{EqSqw}-\ref{EqLor}), are shown in  figure \ref{FigQdisp}. Clearly, both quantities are strongly dependent on $q$ along the $l$ direction but not along the $h$ direction. Such a behavior of $\Gamma_{\bf q}$ have not been reported in other heavy-fermion compounds and will be discussed in section \ref{SecDisc}. 

\subsection{$T$ dependence}
\label{SecTdep}

\begin{figure} 
\begin{center}
\includegraphics[width=7cm]{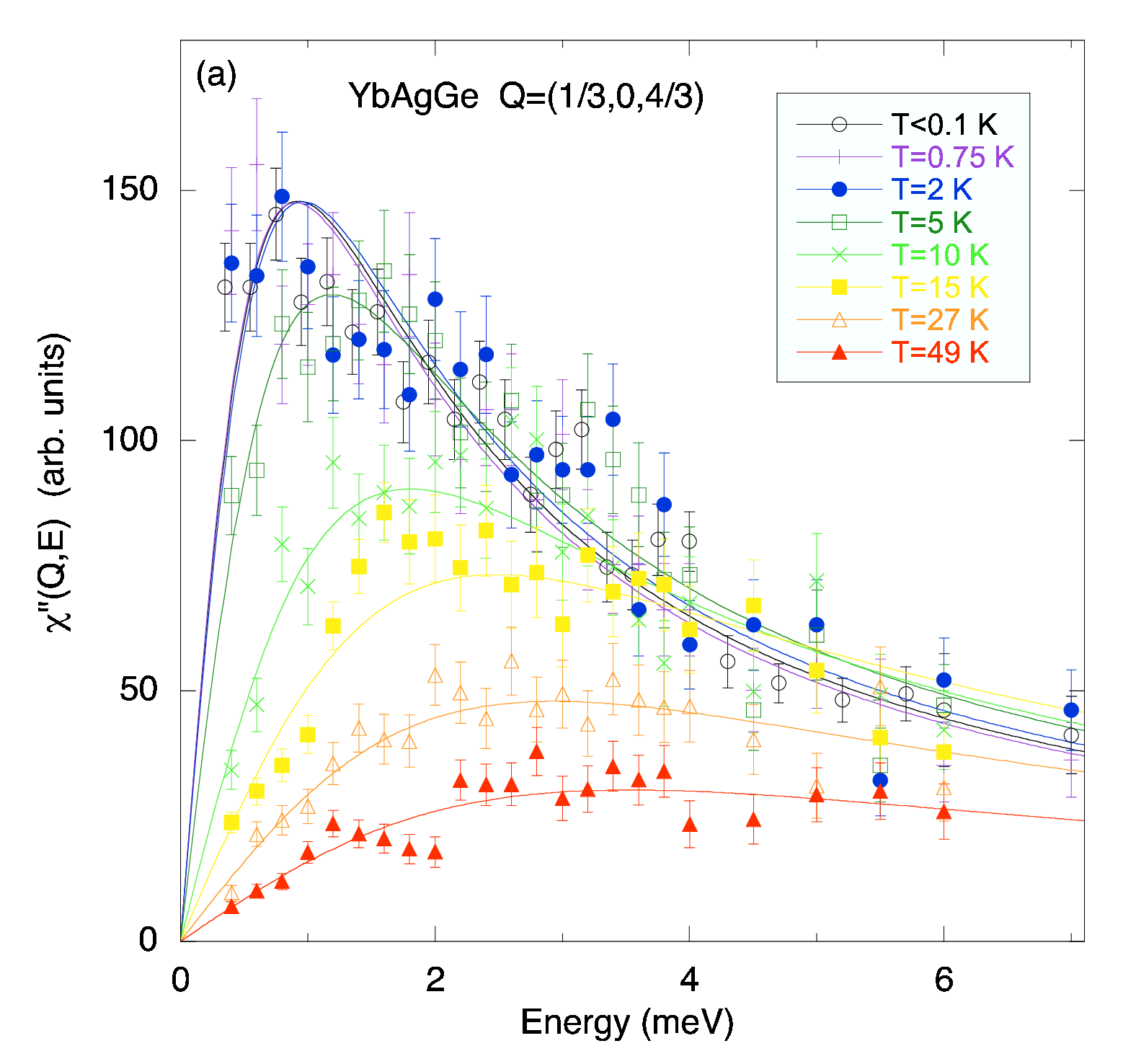}
\includegraphics[width=7cm]{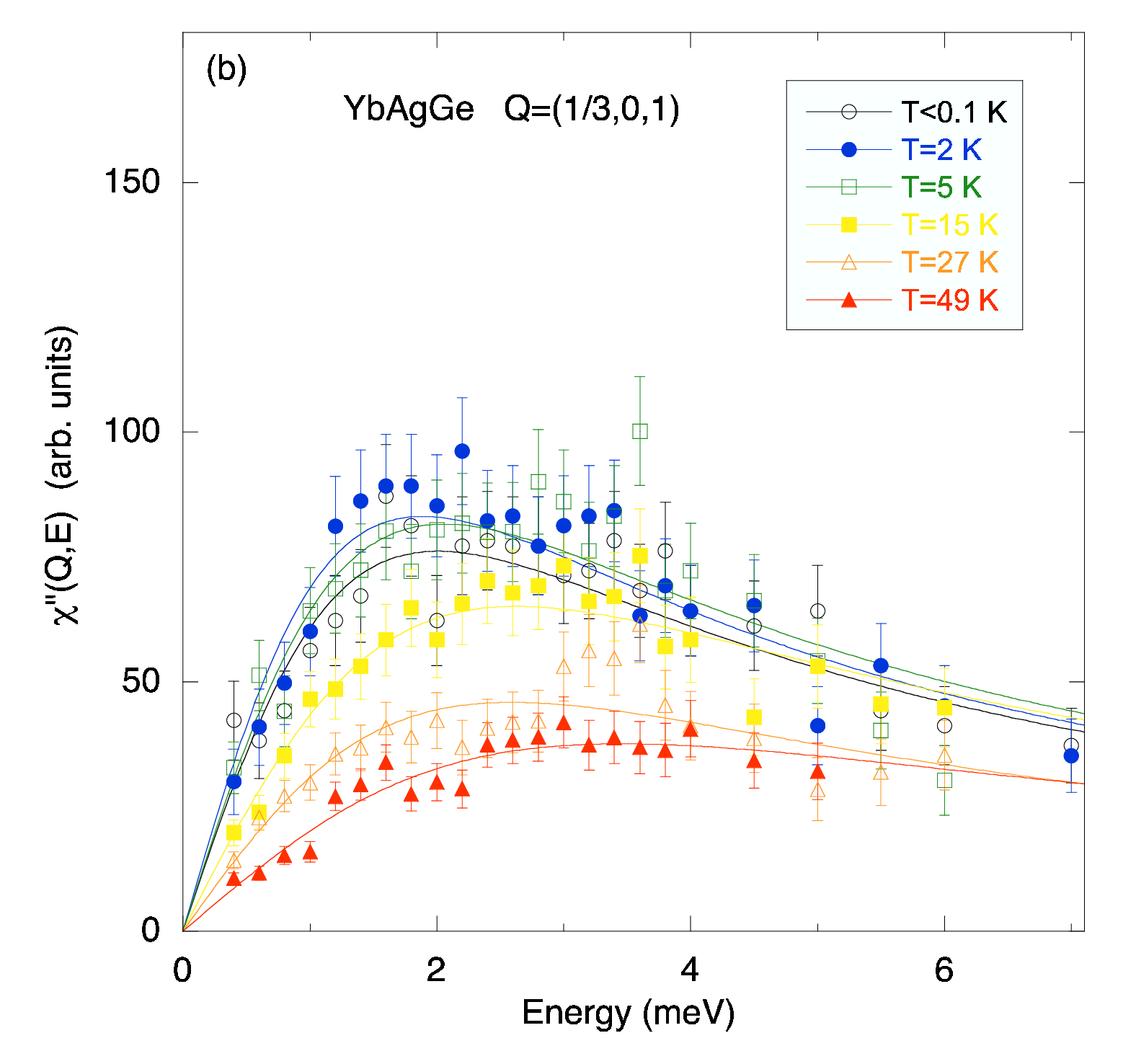}
\end{center}
\caption{Dynamic magnetic susceptibility for different temperatures at (a) the AFM zone center {\bf Q}=(1/3,0,4/3) and (b) the AFM zone boundary {\bf Q}=(1/3,0,1), measured on IN14 using $k_f=1.30$ \Ang. The lines are fits to a quasielastic Lorentzian, Eq.\ (\ref{EqLor}).}
\label{FigEscanHT}
\end{figure}

\begin{figure} 
\begin{center}
\includegraphics[width=10cm]{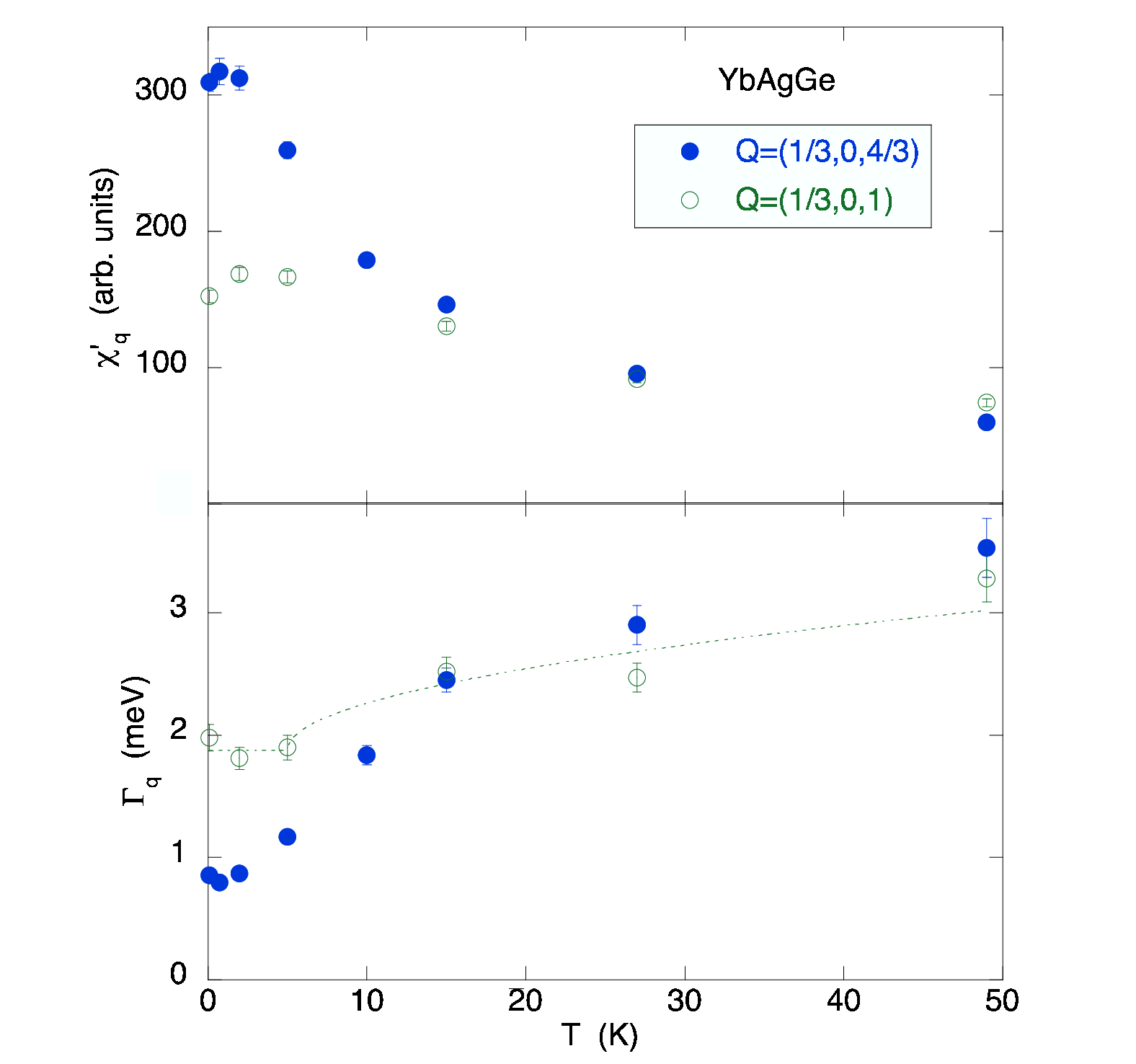}
\end{center}
\caption{Temperature dependence of the static susceptibility $\chi'_{\bf q}(T)$ and the characteristic energy $\Gamma_{\bf q}(T)$ of the quasielastic magnetic scattering at the AFM zone center {\bf Q}=(1/3,0,4/3) (solid circles) and at the AFM zone boundary {\bf Q}=(1/3,0,1) (open circles). The dashed line shows a fit of Eq.\ (\ref{EqGamma}) to the zone boundary energy $\Gamma_0$.}
\label{FigResults}
\end{figure}

The temperature dependence of the dynamic magnetic susceptibility was studied for two wave vectors:  at the antiferromagnetic zone center {\bf Q}$_1$=(1/3,0,4/3)=(0,0,1)$\!+\!{\bf k}$ and at the AFM zone boundary {\bf Q}$_0$=(1/3,0,1). 
These data, shown in figure \ref{FigEscanHT}, were fitted by a quasielastic Lorentzian, Eq.\ (\ref{EqLor}). 
The extracted temperature dependence of the characteristic energy $\Gamma_{\bf q}(T)$ and the static susceptibility $\chi'_{\bf q}(T)$ are shown in  figure \ref{FigResults}.  
The characteristic energy at the AFM zone boundary, $\Gamma_0$, is constant at low temperatures and follows  the standard $\sqrt T$ relation for heavy-fermion materials at higher temperatures, i.e.  
\begin{equation}
\Gamma_0(T)= \Gamma_0(0) + \Theta(T-T^\star)\,A \sqrt T
\label{EqGamma}
\end{equation}
with $\Gamma_0(0)=1.87\ (7)$ meV, $A=0.17\ (2)$ meV\,K$^{-1/2}$, and $T^\star=5.0\ (2)$ K.  Here, $T^\star$ that occurs in the step function $\Theta$ is close to the temperature where the bulk magnetic susceptibility $\chi_a$ peaks \cite{Budko04}. 
The characteristic energy at the AFM zone center, $\Gamma_1$, increases much more rapidly at low temperatures and joins $\Gamma_0$ at $T\approx 15$ K, i.e. in the vicinity of $T_K\sim 20$ K. Above $T_K$, it follows the same $\sqrt T$ relation as $\Gamma_0$. 
The {\bf q}-dependent magnetic susceptibilities $\chi'_{\bf q}$ also show a similar $T$ dependence at
the two wave vectors, with (very approximative) $\chi'\propto 1/\Gamma$ (see section \ref{SecDisc} for a more detailed discussion). 
At temperatures above $T_K\sim 20$ K, the magnetic scattering at the AFM zone center is very similar to that at the zone boundary, showing that the correlations in {\bf q} vanishes at high $T$. 
This is also clearly seen in the {\bf Q}-scans at $E=1$ meV, where the intensity modulation has disappeared at $T=27$ K (see Figure \ref{FigQscan}).  

\section{Discussion}
\label{SecDisc}
As a consequence of the large moment in the doublet crystal-field ground state, the intensity of the low-energy quasielastic magnetic scattering from spin fluctuations in YbAgGe is relatively strong, which allows a detailed analysis of the {\bf q} dependence of the dynamic magnetic susceptibility, \Xqw.  
We find that both \Xq\ and \Gq\ depend not only on the magnitude $q=|{\bf q}|$ of the wave vector  but also on its direction (see  figure \ref{FigQdisp}). 
There are only  few examples of such behavior in $4f$ heavy fermion materials. 
In cerium HF compounds, a {\bf q}-dependent \Xqw\ has been observed in single crystalline samples of 
CeCu$_6$  \cite{Gabe86,Rossa88},  
CeCu$_{5.9}$Au$_{0.1}$ \cite{CeCuAu}, 
CeRu$_2$Si$_2$ \cite{Jacoud89,Regnault90,Kadowaki04},
(Ce$_{0.925}$La$_{0.075}$)Ru$_2$Si$_2$ \cite{Raymond97,Knafo04}, and
CeNi$_2$Ge$_2$  \cite{CeNi2Ge2}. 
However, little is known about the {\bf q} dependence of the {\it characteristic energy} \Gq\ 
in these systems, because of the limited statistics in the neutron scattering measurements. 
An explicit $q$ (but not {\bf q}) dependence of  \Gq\ has only been 
reported in 
CeCu$_6$  \cite{Gabe86} and 
CeRu$_2$Si$_2$ \cite{Jacoud89}.
CeNi$_2$Ge$_2$ is an exception:  the dynamic susceptibility has two components, and for the high-energy ($E=4$ meV) response, \Gq\ is independent of $q$ while  \Xq\  shows a quasi two-dimensional behavior \cite{CeNi2Ge2}. 
Some limited information about the $q$-dependence of \Xqw\  have also been extracted from powder data (examples are 
CeRu$_2$(Si$_{1-x}$Ge$_x$)$_2$ \cite{Rainford92}
and 
Ce(Ru$_{0.24}$Fe$_{0.76}$)$_2$Ge$_2$ \cite{Montfrooij03}), 
but most work assume  that \Xqw\ is either $q$ independent 
or report quantities that are averaged over the  Brillouin zone. 

\begin{figure} 
\begin{center}
\includegraphics[width=14cm]{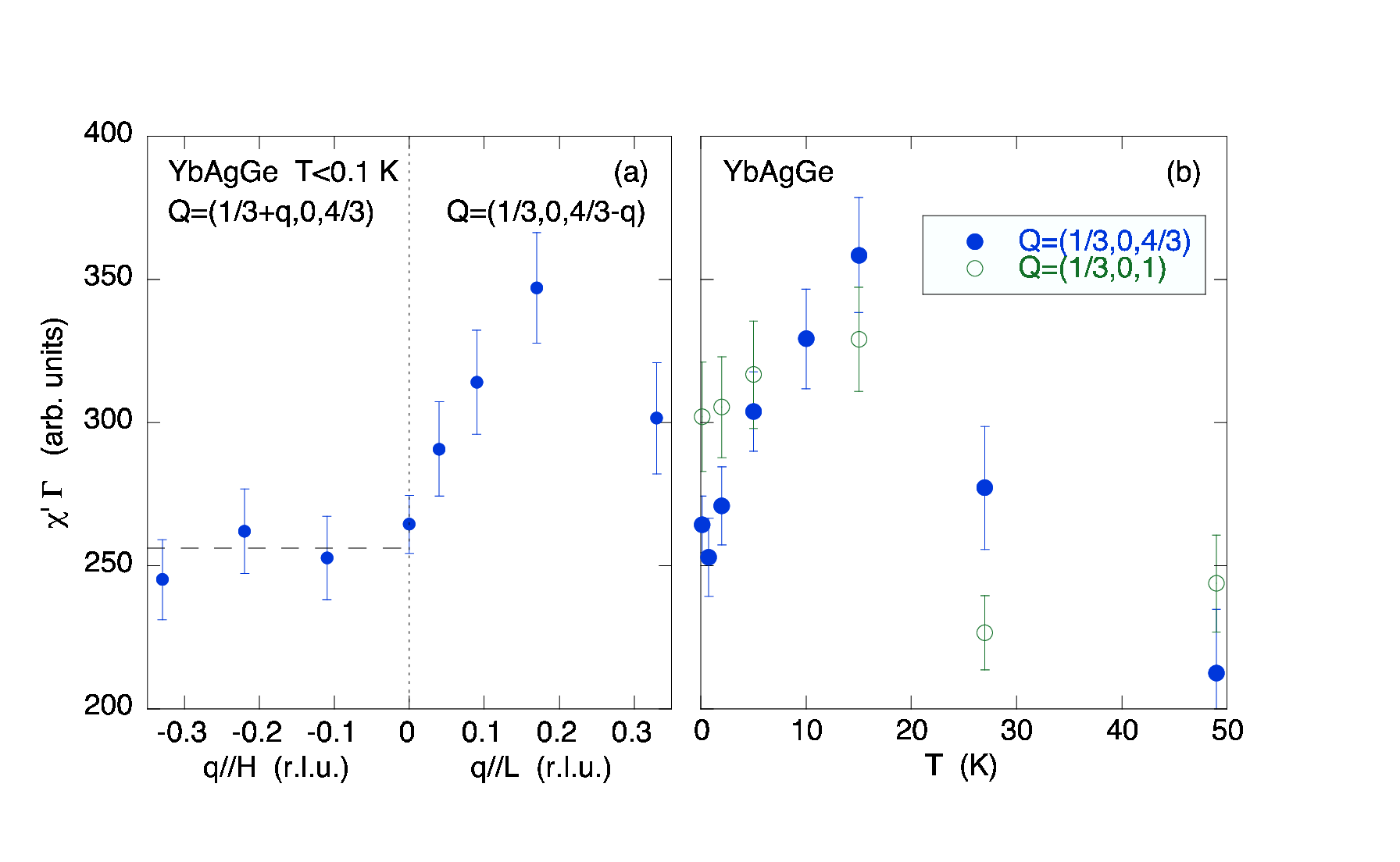}
\end{center}
\caption{The product $\chi'\Gamma$ of the static susceptibility $\chi'_{\bf q}(T)$ 
and the characteristic energy $\Gamma_{\bf q}(T)$ 
of the quasielastic magnetic scattering. 
(a)  $\chi'\Gamma$ as a function of {\bf q} along $h$ (left part) and $l$ (right part).
(b) $\chi'\Gamma$ as a function of temperature for ${\bf q}$ at the antiferromagnetic zone center and at the zone boundary. }
\label{FigXG}
\end{figure}

In spin-fluctuation theories \cite{Moriya95}, the presence of an exchange interaction, 
$J({\bf q})$, between localized moments  described by the local dynamic susceptibility 
$\chi_L(\omega)=\chi'_L/(1-i\omega/\Gamma_L)$, can be described by an RPA-like expression 
$\chi^{-1}({\bf q},\omega)=\chi^{-1}_L(\omega) - J({\bf q})$, 
which leads to  a quasielastic Lorentzian [Eq.\ (\ref{EqLor})]  with
$\chi'_{\bf q}=\chi'_L/[1-J({\bf q})\chi'_L]$ and 
$\Gamma_{\bf q}=\Gamma_L[1-J({\bf q})\chi'_L]$.
It then follows that the product 
$\chi'_{\bf q}\Gamma_{\bf q}=\chi'_L\Gamma_L$ 
becomes $q$ independent, and also temperature independent \cite{Kuramoto87}. 
Thus, in these models, the behavior of $\Gamma_{\bf q}$ follows simply from that of $\chi'_{\bf q}$. 
Experimentally, this relation has been approximately verified in CeCu$_6$  \cite{Gabe86} and CeRu$_2$Si$_2$ \cite{Jacoud89,Kadowaki04}. 
In YbAgGe at low temperatures,  the product $\chi'_{\bf q}\Gamma_{\bf q}$ is constant only in the $h$ direction, where no $q$ dependence of \Xqw\ is observed, while it varies with $q$ in the ``dispersive'' $l$ direction (see figure \ref{FigXG}a). 
Also, the product $\chi'_{\bf q}(T)\Gamma_{\bf q}(T)$
is not constant as a function of temperature,  cf.\ figure \ref{FigXG}b. 

As said, the most characteristic feature of the spin dynamics in YbAgGe is that \Gq\ depends on the direction of the wave vector  {\bf q}. 
The strong variation of \Gq\ and \Xq\ for {\bf q} along the hexagonal $c$ axis is due to antiferromagnetic correlations. 
Similar antiferromagnetic correlations are also expected in the basal plane. 
However, neither \Gq\ nor \Xq\ depend on $q$ for {\bf q} in the basal plane. 
A likely origin for this quasi-one-dimensional behavior 
is the geometrical frustration arising from the triangular coordination of the Yb ions, 
which lie on a quasi-Kagom\'e lattice.
Such frustration is expected to average out the correlations in the basal plane. 
However, it cannot be excluded that the quasi-one-dimensional behavior arises from Fermi surface effects. 
In addition to the particular {\bf q} dependence of \Gq\ and \Xq, we also find that the spins fluctuate predominantly in the basal plane, due to the anisotropy of the crystal-field interactions.

With increasing temperature, the characteristic energy of YbAgGe increases much faster at the magnetic zone center than at the magnetic zone boundary (see  figure \ref{FigResults}). Above $T_K\approx 20$ K, the modulations in {\bf q} disappear and all $\Gamma_{\bf q}$ become identical. In this temperature region,  $\Gamma_{\bf q}$ follows the $\sqrt T$ dependence characteristic for heavy-fermion compounds. 
Similar behavior has been found in other heavy-fermion systems, such as
CeCu$_6$ \cite{Rossa88},
CeRu$_2$Si$_2$ \cite{Regnault90}, 
(Ce$_{0.925}$La$_{0.075}$)Ru$_2$Si$_2$ \cite{Knafo04}, and
Ce(Ru$_{0.24}$Fe$_{0.76}$)$_2$Ge$_2$ \cite{Montfrooij03}, where
$\Gamma_{{\bf q}={\bf k}}$ at the critical wave vector {\bf k} joins from below 
$\Gamma_{{\bf q}\neq{\bf k}}$ at temperatures of the order of the Kondo temperature $T_K$. 

For a system close to a quantum critical point, 
the dynamic susceptibility is expected to follow so-called  $\omega/T^\beta$ scaling, 
\begin{equation}
\chi''({\bf q},\omega)T^\alpha=f(\omega/T^\beta),
\label{EqScaling}
\end{equation}
where $f$ is a scaling function, $\alpha\sim 1$, and $\beta=$ 1--1.5 \cite{Coleman01}. 
This implies that the only energy scale in the system is the temperature, 
and in particular, that $\Gamma_{\bf q}(T)\rightarrow 0$ as $T\rightarrow 0$. 
This is clearly not the case for YbAgGe in zero magnetic field, where $\Gamma_{\bf q}$ is finite at all 
{\bf q} in the limit $T\rightarrow 0$. YbAgGe in zero field is thus a well-behaved heavy-fermion without $\omega/T^\beta$ scaling. 

The next step in our investigation of single-crystalline YbAgGe is to study by  inelastic neutron scattering the evolution of the low-temperature dynamic magnetic susceptibility as the system is tuned by an applied magnetic field from the Fermi-liquid regime to the non-Fermi-liquid regime, where $\omega/T^\beta$ scaling might be observed.

\ack We thank X. Tonon, R. Down, and A. Church for cryogenic support. Institut Laue-Langevin
(Grenoble, France) provided the neutron beam time on IN14. 
Ames Laboratory is operated for the U.S. Department of Energy by Iowa State University under Contract No. W-7405-ENG-82.  This work was in part supported by the Director for Energy Research, Office of Basic Energy Sciences.

\end{document}